\documentclass[3p,number]{elsarticle}%  
\usepackage{pifont}
\usepackage{natbib}
\usepackage[utf8]{inputenc}
\usepackage[T1]{fontenc}
\usepackage{amssymb}
\usepackage{amsmath}
\usepackage{amsfonts}
\usepackage{xifthen}
\usepackage{graphicx}
\usepackage{enumerate}
\usepackage{framed}
\usepackage{tikz}
\usetikzlibrary{petri}
\usepackage{hyperref}
\newcommand{\comment}[1]{}

\newcommand{\stufforempty}[1]{%
  \ifthenelse{\isempty{#1}}{\;}{#1}}

\newtheorem{theorem}{Theorem}

\newtheorem{proposition}[theorem]{Proposition}

\journal{Theoretical Computer Science C}
\begin{document}
\begin{frontmatter}
\title{Spectral concepts in genome informational analysis}
\author[vb]{V.~Bonnici}
\ead{vincenzo.bonnici@univr.it}
\author[gf]{G.~Franco}
\ead{giuditta.franco@univr.it}
\author[vm]{V.~Manca\corref{cor}}
\ead{vincenzo.manca@univr.it}
\cortext[cor]{Corresponding author}
\address[vb,gf,vm]{University of Verona, 
Department of Computer Science, 
Strada Le Grazie, 15 -- 37134 Verona, Italy}
\begin{abstract}
The concept of $k$-spectrum for genomes is here investigated as a basic tool to analyze genomes. Related spectral notions based on $k$-mers are introduced with some related mathematical properties which are relevant for informational analysis of genomes. 
Procedures to generate spectral segmentations of genomes are provided and are tested (under several values of length $k$ for $k$-mers) on cases of real genomes, such as some human chromosomes and Saccharomyces cervisiae.
\end{abstract}
\end{frontmatter}

\section{Introduction}
The notion of spectrum comes from physics, usually related to 
a function of frequency (namely frequency spectrum, or power spectrum of a signal). The term applies to any 
 signal decomposed along a variable, such as energy in electron spectroscopy or mass-to-charge ratio in mass spectrometry. In a wider sense, a spectrum is a distribution associated to a phenomenon, where a measure (intensity, density, frequency) is associated to each value taken from a variable. In fact, spectroscopy originated as the study of light intensity at the different wavelength of its components. 
 
A linearized genome (that is, observed by neglecting the secondary and tertiary structure of the double helix) is a long sequence of characters from the DNA alphabet $\{A, T, C, G\} $, which may be seen as an information source, composed by a finite set of signals (e.g., words) emitted (or occurring) at a given ``frequency''. In particular, if we restrict ourself to consider all genomic substrings of a fixed length $k$ (factors of length $k$, here called $k$-mers, in other contexts also called $k$-grams), we may call {\it $k$-spectrum of a genome} the distribution which associates to each $k$-mer the (normalized) number of times that it occurs in the genome (that is, its multiplicity, which is called frequency when normalized). If we denote by $D_k(\mathbb{G})$ the set of $k$-mers occurring in a genome $\mathbb{G}$, and by $mult_\mathbb{G}(\alpha)$ the multiplicity of a $k$-mer $\alpha$ in  $\mathbb{G}$, the {\it k-spectrum} of a genome $\mathbb{G}$, is the multiset $$spec_k(\mathbb{G}) = \{(\alpha,  mult_\mathbb{G}(\alpha)) \; | \; \alpha \in D_k(\mathbb{G}) \}.$$
In sequence analysis, the term ``spectrum'' is used in many contexts and with several meanings (see for example, \cite{Spt1,Spt0,Spt2, Coded, error2}), in particular to tackle the problem of reconstructing a string from the compositions of its substrings. Other terms are used with a similar meaning, as {\it bags of words}, where a $k$-spectrum is represented by a multiplicity vector, where the $i$-component gives the multiplicity of the $k$-mer of position $i$ in some prefixed (for example lexicographic) order~\cite{Lan}. The {\it genomic profile} is instead a related concept with a different meaning, as it accounts for the distribution of quantity of $k$-mers over their multiplicity, by ignoring the information on the specific $k$-mer sequences.

In general, string reconstruction refers to the long standing problem of recovering a string based on some information about its substrings (i.e, sequences composed of consecutive elements of the string) or subsequences (i.e, sequences composed of possibly non-consecutive elements of the string)~\cite{Scott}. Above problems may be formulated under specific hypotheses, for example whether or not we are given the order of the substrings in the string, or the order of the bits in each substring~\cite{bello}, also, predefined constraints may be imposed on the properties of the strings one seeks to reconstruct (see coded strings in~\cite{Coded} or~\cite{MJ} for a model of DNA self-assembly). The first instance of a coded sequence reconstruction problem was studied by Levenshtein~\cite{Lev}, who posed the sequence reconstruction problem for strings drawn from an error-correcting codebook (setup in which not all substrings in the $k$-spectrum are received). Reconstruction of a string from {\it a few} substrings was considered in~\cite{traces1,Spt1}  as well as the case where, for each substring, only the composition is given, neither the
order of the symbols within it nor the substring’s location in the original string. For example, in~\cite{error2} reconstruction of strings based upon their error-prone substrings spectrum is developed, while the noisy setup of this problem was first studied by Gabrys and Milenkovic~\cite{Coded}.

In several works the initial information is given in terms of the sequence spectrum, over a binary alphabet, comprising all distinct substrings (of variable length) appearing in the string: the multiset spectrum~\cite{Spt2}, comprising the multiset of the substrings of the string; the $n$-deck~\cite{Lev}, comprising all subsequences of the string of length
$n$; sequence traces~\cite{traces2}, corresponding to randomly selected subsequences of the
string; or multiset compositions~\cite{bello}, providing information about the composition of substrings only. These approaches are often motivated by molecular biology problems: mass-spectrometry protein sequencing~\cite{error1,bello}, DNA-based data storage, where
the strings to be sequenced are user-defined and synthetically generated, and hence allowed to have arbitrary content~\cite{Lev}, and account for types of errors  that are encountered in both DNA synthesis and sequencing~\cite{error1}.  For example, the nanopore sequencing technology (developed by Oxford university) motivated the assumption of ``traces'', that are genomic subsequences long $k$ which have passed  through $k$ deletion channels, deleting each character of the original string with a given probability.

In our (above defined) notion of genomic $k$-spectrum we are given {\it all} the {\it error-free} $k$-mers of the genome, with their multiplicity value, whereas we do not know their order on the genome nor their exact position. 
We focus on the relevance of the $k$-spectrum for genomes analysis,
by exploiting the information contained in a $k$-spectrum to define new spectral notions 
%{\it spectral segments} and  {\it spectral segmentations}, 
which allow compact representations of genomes.

Many concepts based on $k$-mers, or analogous notions of {\it scattered factors} and  {\it de Bruijn sequences}, are shared by many fields in Computer Science, e.g. combinatorics on words, frequently with different names and aims~\cite{BK,MA00,G}. However, in this paper, $k$-mers and related spectra are mainly considered in the perspective of their possible elongation within the genomes, a feature which turns out very significant for their enormous lengths (from millions to billions of characters).
%So, I would like to read in this paper, some remarks showing how k-spectrum is related to
%concepts as Scattered Factors, de Bruijn sequences, composition of k-mers (see [G] below).
%More specifically, it seems to me that k-spectrum is nothing but a formal polynomial of degree k [see BK, page 44-45 for example; or other classical books on
%combinatorics on words]
%or in [MA00], see below.
%
Apparently  related to the aim of this paper there are also the articles~\cite{bello} and~\cite{error1}, motivated by mass-spectrometry protein sequencing. They report the simply-stated problem of reconstructing a string from the multiset of its substring compositions. However, the assumed substring composition multiset contains substrings of variable lenghts, and the interesting results in~\cite{bello} are found only for short strings (e.g., strings of length 7 can be reconstructed uniquely up to reversal), while the general problem formulation being a combinatorial simplification of the well-known turnpike problem. Interestingly, general classes of strings that cannot be distinguished from their substring compositions are provided in~\cite{error1}. 

Works dealing with (real) genome recombinations and reconstructions often assume to start with a set of substrings for which multiplicity is not known, as it is the case for ciliates rearrangements~\cite{jonoska, jonoska2}. On the other hand, current methods for DNA assembling which belong to “next-generation” (of second, third, and fourth generation) sequencing (NGS) allow us to have several duplicates of given $k$-mers.

In this paper we aim at providing correspondences between informational profiles and biological interpretations, as in~\cite{relevant,relevant2}, where genomic sequence entropy and complexity are investigated to discover evolution patterns, also discussed in~\cite{manca-marvelous-2018}. Some concepts related to genomic $k$-spectra have been introduced in our previous work~\cite{castellini2012dictionary,sec14,infogenomics-elsevier-2015,manca-cmc16,manca-tcs-2017,RDD-2015,igtools,laws,random-strings,manca-cmc16,sec14,recent} and are here recalled in Section 2. In particular, the distinction between {\it hapax} and {\it repeat} substring, as defined later on, and all the notions related to them~\cite{Mil1,Mil2}. Their importance in genome analyses emerged in the course of investigations developed by means of information theory concepts. After a brief discussion  in Section 2 on genomic $k$-spectra, in Section 3 we introduce related notions of {\it spectral segments}, {\it spectral segmentation}, {\it k-univocal genomes} and {\it genome spectrality}, by showing their relevance to recognize long portions of genomes univocally identified by spectra. Section 4 concludes our contribution with a few final comments and observations.

\section{Properties related to $k$-spectra}
In this section we present some informational properties of a genome $\mathbb{G}$, which are given within its $k$-spectrum: $$spec_k(\mathbb{G}) = \{(\alpha,  mult_\mathbb{G}(\alpha)) \; | \; \alpha \in D_k(\mathbb{G}) \}.$$
The usual operations of multiset-sum and  multiset-difference  are assumed for $k$-spectra.  In particular, the elements of a multiset-sum $A+B$ are those of $A$ or $B$ with the multiplicities given by the sum of their multiplicities in $A$ and $B$, while those of $A-B$ have as multiplicities the differences of multiplicitiies, by setting to zero the negative values. Moreover, a multiset does not change if a new element with zero multiplicity is added or removed, therefore a multiset having only zero multiplicities can be considered equivalent to the empty  set.

The support of a $k$-spectrum of a genome $\mathbb{G}$ is the {\it k-dictionary}~$D_k(\mathbb{G})$, that is, the set of all $k$-mers occurring at least once in $\mathbb{G}$. Notice that we are interested to study this specific dictionary with a prefixed $k$, rather than the dictionary of genomic words $$D(\mathbb{G}) = \bigcup_{1 \leq k \leq |\mathbb{G}|} D_k(\mathbb{G}).$$
However, all the notions introduced here can be extended to this dictionary $D$, by replacing $spec_k(\mathbb{G})$ by $spec_D(\mathbb{G})$:
$$spec_D(\mathbb{G}) = \{ (\alpha,  mult_\mathbb{G}(\alpha)) \; | \; \alpha \in D \}.$$

Here we point out that reversal strings have $k$-spectra with reversal words and identical multiplicities, and two strings with the same $k$-spectra have the same length. We indicate by $|\mathbb{G}|$ the length of  $\mathbb{G}$. The count of possible positions for $k$-mers on the genome $\mathbb{G}$ is $ |\mathbb{G}| - k +1$, which corresponds to the sum of the multiplicities (over $D_k(\mathbb{G})$) present in the $k$-spectrum.

A $k$-mer $\alpha$ such that $mult_\mathbb{G}(\alpha)) = 1$ in the literature is called a {\it hapax} (plural hapaxes) or unique string, while $\alpha$ is called a {\it repeat} if it occurs many times, that is, $mult_\mathbb{G}(\alpha) > 1$. Then, the $k$-dictionary of a genome may be decomposed in two disjoint subsets, the set $H_k(\mathbb{G})$ of hapaxes and its complementary set $R_k(\mathbb{G})$ of repeats, by omitting $(\mathbb{G})$ when the genome is clear from the context.

For example, in the genome $abcbbabcaa$ substrings  $ab, bc$ are repeats occurring two times; all the substrings shorter than 2 are repeats; the substring $abc$ is a maximal repeat, because it occurs two times and no longer repeat occurs in the genome; the substrings $cb, bb, ba, ca, aa$ are hapaxes because each of them occurs only once in the genome; all the substrings longer than 3 are hapaxes, moreover $cb, bb, ba, ca, aa$  are also minimal hapaxes, because no hapax shorter than them occurs in the genome.

If a genomic string includes a $k$-hapax as a substring, for some $k$, it is a hapax too, 
while if it is a substring of a $k$-repeat, for some $k$, it is a repeat too.
%while if it includes a $k$-repeat, for some $k$, it is a repeat.

Given a genomic $k$-spectum, some informational indexes over the genome may be defined~\cite{castellini2012dictionary,manca-infobiotics,infogenomics-elsevier-2015,manca-cmc16,laws,manca-tcs-2017}, which have nice mathematical properties, as in the following. The average multiplicity of $k$-mers $LX_k(\mathbb{G})$ is called $k$-{\it lexical index}, the {\it maximal repeat length} is denoted by $mrl(\mathbb{G})$, the {\it minimal hapax length} is denoted by $mhl(\mathbb{G})$. For example, in the genome considered above, we have that $mrl(abcbbabcaa) = 3$ and $mhl(abcbbabcaa) = 2$.
By definition, any genomic string longer than its $mrl$ is a hapax, and shorter than its $mhl$ is a repeat.

\begin{proposition}
Maximal repeats over genomes have multiplicity equal or less than 5.
\end{proposition}
{\bf Proof} If $\alpha$ is a maximal repeat, none of its extensions $\alpha x$ with $x \in \{A, C, G, T\}$ can occur twice, otherwise $\alpha$ would not be maximal. This means that symbols $x$ after the occurrences of $\alpha$ in the genome have to be different, that is $\alpha$ may occur at most four times in the middle of the genome, and additionally
$\alpha$ may occur as a suffix of the genome. In conclusion, a maximal repeat  can have at most five possibilities of occurring.~\qed

An interesting speculation to study genomic strings may be developed around the inclusion of all $k$ long words in the $k$-dictionary, or some possible absence of such words. To this purpose, over a genome $\mathbb{G}$, we define the {\it maximal complete length} $mcl(\mathbb{G})$ as the maximal $k$ such that all possible $k$ long strings occur in $\mathbb{G}$), and the {\it minimal forbidden length} $mfl(\mathbb{G})$ as the minimal length $k$ such that at least one $k$-mer does not belong to the genomic $k$-dictionary (in the literature this is called $k$-forbidden word or $k$-nullomer). 

By definition, thus, in any genome, $mcl = mfl -1$. Moreover, it holds the following

\begin{proposition}
In all genomes $mfl \leq mhl +1$.
\end{proposition}
{\bf Proof} Any hapax $\alpha$ is followed over the genome by only one symbol of the alphabet. Therefore, if $|\alpha| = mhl$ then we may find three forbidden words long $mhl +1$.~\qed

\

Since $mcl \leq mhl$ (as a consequence of the observations above), any genomic string shorter than the $mcl$ is  a repeat. If we call $LG = \lg_4(|\mathbb{G}|)$, the {\it logarithmic length} of $\mathbb{G}$, we have also the following property for the $mcl$ index:

\begin{proposition}
In any genome $\mathbb{G}$ the inequality $mcl < LG$ holds.
\end{proposition}
{\bf Proof} 
Let $n$ be the length of genome $\mathbb{G}$. Since, by definition of $mcl$, we have $4^{mcl} \leq n - k +1$, then $mcl \leq \lg_4(n - k +1) < LG .$~\qed

\

A genome $\mathbb{G}$ is a $k$-{\it hapax genome} when its dictionary $D_k(\mathbb{G})$ consists only of hapaxes, it is also a $k$-{\it complete} genome if $D_k(\mathbb{G})$ consists of all $4^k$ $k$-mers (in this case the length of $\mathbb{G}$ is
 $4^k + k -1$).

The toy genome $aaccggttagatctg$ is a $2$-hapax genome,  while $aaccggttagatctgca$ is a $2$-hapax complete genome. It is complete because all pairs occur and it is hapax because it has the minimum length $17 = 2^4 +2 -1$ capable of containing all the possible pairs. However, if we permute the last two symbols the result is not anymore a hapax genome: $aaccggttagatctgac .$

Given the $k$-spectrum of a genome $\mathbb{G}$, the $k$-{\it entropy}  $E_k(\mathbb{G})$ is defined as the Shannon entropy of the probability distribution assigning to any $k$-mer its frequency:
$$E_k(\mathbb{G}) = \sum_{\alpha \in D_k(\mathbb{G})} p(\alpha) \log_2 p(\alpha) \quad \mbox{where} \quad p(\alpha) = \frac{mult_\mathbb{G}(\alpha)}{|\mathbb{G}| - k +1}$$

According to the equipartition property of Entropy \cite{shannon,laws} we may deduce the following result:
\begin{proposition}
Any $k$-hapax genome $\mathbb{G}$ of length $n$ has the maximum value of $E_k(\mathbb{G})$ among all genomes of the same length.
\end{proposition}

We conclude this section with the observation that the notion of $k$-spectrum can be extended to any genomic distribution, that is, to any variable $X$ associated to a genome and assuming a (finite) set $A$ of values, by considering the number of times any value is assumed by $X$.  In this sense $X$ can be considered an information source, as defined within Shannon theory~\cite{shannon}, therefore the informational nature of genomic spectra clearly appears as a powerful feature to analyse and compare genomes.

\section{Spectral segmentation of genomes}
In this section some notions based on genomic spectra are introduced that shed a new light
on the structure of genomes.

Intuitively, we aim at finding long genomic segments constructed by overlap concatenation of $k$-mers (out of the $k$-spectrum, then coming with their multiplicity), with an overlapping long $k-1$. Namely, we look for {\it unary paths} (paths of nodes with degree 1) in de Brujin graph (essential tools in genome sequencing \cite{compeau}) under the constraint given by the multiplicity: each $k$-mer node may be passed through only the number of times given by the $k$-mer multiplicity in the $k$-spectrum.

Two $k$-mers ($\alpha$, $\beta$) may be {\it $k$-concatenated} if the $(k-1)$-long suffix of $\alpha$ equals the $(k-1)$-long prefix of $\beta$, that is, there exist $\gamma$ such that $\alpha = x \gamma$ and $\beta = \gamma y$, with $x$ and $y$ being symbols from the alphabet. The concatenation results in $\alpha \cdot_k \beta = x \gamma y$, 
which is the string $\alpha y$ obtained by right elongation of the string $\alpha$ with the symbol $y$ (suffix of $\beta$).

In a $k$-spectrum one may find couples of strings which may be $k$-concatenated, either according to the way they are $k$-concatenated over the genome or differently, and this process may be iterated, to assemble either the genome or other different strings. Even under the constraint given by the multiplicity, to use any $k$-mer only the number of times allowed by its multiplicity, strings that do not occur in the genome may be obtained.

For example, the two different strings: $G = aaggccgaagggcacccaa$ and $G' = aagggcacccaaggccgaa$ (the second being obtained by the first one by swapping the prefix $aaggccgaa$ of $G$ with the suffix $aa$ of $G^\prime$) share the following  2-spectrum:
$$\{(aa, 3) , (ag, 2) , (gg, 3) , (gc, 2)  ,  (cc, 3),  (cg, 1) , (ga, 1) , (ca, 2) , (ac , 1)\}.$$
These two different strings  are obtained as in the following, as a result on the right of $\to$, by iterating the $2$-concatenation in the sequences reported on the left of $\to$, where the same 2-mers occur, with the multiplicities of their common 2-spectrum, but with different orders:
$$aa, ag, gg, gc, cc, cg, ga, aa, ag, gg, gg, gc, ca, ac, cc, cc, ca, aa \to aaggccgaagggcacccaa$$ 
$$aa, ag,  gg, gg, gc, ca, ac, cc, cc, ca, aa, ag, gg, gc, cc, cg, ga, aa \to aagggcacccaaggccgaa .$$

Since the $k$-spectrum of a genome in general does not identify it in a univocal way, we say {\it k-univocal} (in spectrum) a genome $\mathbb{G}$ if no genome different from $\mathbb{G}$ exists that has the same $k$-spectrum of $\mathbb{G}$. 
Of course, a $k$-univocal genome may always be obtained for suitably high values of $k$ 
(two trivial cases being the length of the genome itself and mrl + 2, since $k$-mers and their possible overlapping are all hapaxes). Given a genome  $\mathbb{G}$, it would be interesting to know which is the minimal $k$ for which it is possible a unique reconstruction from the $k$-spectrum by $k$-concatenation. 
The following discussion and toy examples of $k$-spectra may help to familiarize with these concepts.

Strings $a^3 b^2 a^5$ and $a^5 b^2 a^3$ have the same 2-spectrum: $\{(aa, 6), (ab, 1),$ $(bb, 1), (ba, 1)\}$, then they are not 2-univocal, but they are 5-univocal. In particular, they are both 5-hapax, with $$spec_5(a^3 b^2 a^5) \not =  spec_5(a^5 b^2 a^3) $$ since $aaaab$ occurs only in $a^5b^2a^3$ while $baaaa$ occurs only in $a^3b^2a^5$. However, {\it a $k$-hapax genome is not necessarily $k$-univocal}, as it may easily seen from the sequences $aabcac$ and $acaabc$ having both the 2-spectrum $\{(aa, 1), (ab, 1), (bc, 1), (ca, 1), (ac, 1)\}$.

If $\mathbb{G}$ is not $k$-univocal, by definition there is at least one different  genome $\mathbb{G'}$ having the same spectrum of $\mathbb{G}$. This means that at least some of the $k$-mers of the spectrum occur in the two genomes $\mathbb{G}$ and $\mathbb{G'}$ with different  reciprocal orders. 
Also, from the examples above, we may deduce that two (different) genomes with the same $k$-spectrum may be obtained by $k$-concatenation of the $k$-mers in a different order. 

An interesting question here is: which are the longest arrangements of $k$-mers having, in both genomes $\mathbb{G}$ and $\mathbb{G'}$, the same $k$-mers occurrences with the same order? In an extreme case these arrangements coincide with the $k$-mers, however in general these portions can be longer than $k$, as we will see in the following subsection, 
where spectral maximal segments are defined. It could be interesting to find such common components occurring in genomes having the same $k$-spectrum.

\subsection{Spectral segments}
 
Given a $k$-spectrum H, we call $k$-{\it spectral segment} an iterated $k$-con\-ca\-te\-na\-tion of $k$-mer   %
copies in H. Being each $k$-mer present in H with a multiplicity (number of copies), all the $k$-{spectral segments} that can be formed have to involve $k$-mers not more times than the number of their occurrences in H. 
This process is like having bricks in a limited number of copies, and compose them until they may be used up.

A $k$-spectral segment may result in a string which is not a substring of any genome having H as a $k$-spectrum. For example, the string $canegattogallogattone$ has the $4$-spectral segment $canegattone$ which is not a substring occurring in the string. This feature is due to the possibility to have more than one way to elongate a $k$-mer out of the $k$-spectrum.

Let us here point out that this phenomenon is present also for $k$-hapax genomes. As an instance, see genome $abcabbaba$, with 3-spectrum: $$\{(abc,1), (bca,1), (cab,1), (abb,1) (bba,1), (bab,1) (aba,1)\}.$$ In this spectrum where the 3-spectral segment $abbabca$ can be obtained by iterating 3-concatenation over the following sequence on the left of $\to$: 
$$(abb, bba, bab, abc, bca) \to abbabca$$
however the obtained string (on the right of $\to$) is not a substring of the genome $abcabbaba$, which in turn can be obtained by concatenating in a different way the hapax 3-mers of the spectrum:
$$(abc , bca, cab,  abb,  bba,  bab, aba) \to abcabbaba .$$
These examples  (the last one and of genome $canegattogallogattone$) show an important not so obvious aspect of $k$-spectral segments.
%This is due to the possibility to obtain two different 3-concatenations, on the two 3-hapaxes with suffix $ab$ and the two 3-hapaxes $abc$ and $abb$. 

Of course, if a $k$-mer is a hapax by definition it may be univocally elongated {\it over the genome}, but the converse does not hold, because  a $k$-mer may always be elongated by the same symbol (in all its occurrences) without any necessity to be a hapax. 

Furthermore, we notice that in the example above we may ``maximally'' elongate the 3-spectral segment $abcabbaba$ (by $3$-concatenation) and obtain another genome $abbabcaba$ (of the same length).  This segment is maximal because ``we have consumed all the bricks'', that is, the sum of multiplicities in the $k$-spectrum is an upper bound to the number of times we may iterate the $k$-concatenation. 

The capability to obtain $k$-spectral segments which are not genomic substrings depends on the $k$ value as well. Indeed, the above genome $abcabbaba$ (which is of course also 5-hapax) has 5-spectral segments (such as $abbaba$) which are all genomic substrings.

Due to the observations and discussion so far, we need to introduce some helpful more specific definitions. In particular, if we include $k$-mers of the $k$-spectrum in the definition of $k$-spectral segments, we say that a $k$-spectral segment $\alpha$ of H
is {\it univocally elongated} (on the right) in H if the $k$-mer suffix of $\alpha$  can be $k$-concatenated with only one $k$-mer $\beta$ in $H$. Analogously, $\alpha$ is {\it univocally elongated} (on the left) if only one $k$-mer $\beta$ in $H$ can be $k$-concatenated with the $k$-mer prefix of $\alpha$. Moreover, $\alpha$ is {\it maximally univocally elongated} when, after its last unique elongation, on the right or on the left, $\alpha$ cannot be further univocally elongated in H (on the right or on the left). 

We shortly call  {\it  $k$-spectral maximal segments} the maximally univocally elongated $k$-spectral segments (``spectral maximal segments'' or only ``maximal segments'' when $k$-spectrum is clear from the context). They represent genomic portions that are blocks common to all the genomes having the same $k$-spectrum, a sort of signature associated to a genomic $k$-spectrum. In the following, experimental results on some human chromosomes are reported along with seven tables, where the term ``Maximals''  stands for ``Maximal segments'', which have been specifically computed and counted in number.

The $k$-spectral maximal segments are the longest arrangements of $k$-mers having the same $k$-mers occurrences, with the same order, in all genomes with the same $k$-spectrum. We say $k$-{\bf segmentation} of H the multiset of the $k$-spectral maximal segments of H (a maximal segment may occur in H with a multiplicity greater than 1). The following two procedures provide the $k$-segmentation of a $k$-spectrum~H according to different strategies.

%\subsection{Procedures for $k$-segmentations}
\subsection*{Procedure 1 to compute $k$-segmentations} 
The procedure starts by assigning to the current string $w$ one $k$-mer chosen out of $H$ and, at each step, searches in $H$ for the $k$-mer that uniquely elongates $w$, on the right, or on the left,  by updating $H$ after removing  the  $k$-mer occurrence used in the elongation. If more than one possible $k$-mers of $H$ can be concatenated to both the extremities of the current string, then the concatenation process halts, and the string obtained so far is produced in output, as a $k$-spectral maximal segment. Then, the procedure restarts with a similar process with another $k$-mer of $H$.  The procedure halts when $H$ becomes empty.\\ 

In the procedure above it is not essential to establish the choice criterium for the initial string $w$ (the first $k$-mer in the lexicographic order could be one possibility). Some of the $k$-spectral maximal segments produced in output during the whole process above can be generated with a multiplicity greater than 1. The multiset of these segments with their respective multiplicities is the $k$-segmentation of H. Clearly, in the genomes with $k$-spectrum $H$, the $k$-spectral maximal segments of the obtained $k$-segmentation of $H$ occur with different possible orders that are specific of each genome.

If we change perspective, and assume to start from the genomic sequence itself, we may compute its spectral segmentation in a more efficient way. This is indeed the strategy we have followed by procedure 2, in order to compute the $k$-segmentations of real genomes: given a genome $\mathbb G$, a length $k$, and its $k$-dictionary $D_k(G)$, the $k$-spectral maximal segments of $\mathbb G$ are retrieved by mapping {\it univocally elongated} words of $D_k(G)$ at their starting positions in $\mathbb G$. %According to this strategy, the following procedure provides the  $k$-segmentation of $\mathbb G$.\\

\subsection*{Procedure 2 to compute $k$-segmentations}  A $k$-mer $\alpha \in D_k(\mathbb G)$ is {\it univocally elongated} in $G$ if $|\{  \beta \in D_k(G) :  \alpha[2...k] = \beta[1...k-1]  \}| = 1$. A Boolean array $A$, such that $|A|=|\mathbb G|$, is initialized to be {\it false} in every position. Let $pos(\alpha,{\mathbb G})$ to be the set of starting positions in $G$ where $\alpha$ occurs, namely $pos(\alpha, {\mathbb G}) = \{ i : {\mathbb G}[i,...,i+ k -1] = \alpha \}$. For each $\alpha \in D_k(G)$ such that $\alpha$ is {\it univocally elongated} in ${\mathbb G}$, the algorithm sets as $true$ all the positions where an occurrence of $\alpha$ starts in ${\mathbb G}$, namely $A[i] = true$ for each $i \in pos(\alpha,G)$. Then, the $k$-segmentation is retrieved by scanning for consecutive runs of $true$ values in $A$, and a $k$-spectral maximal segment is a substring ${\mathbb G}[i...j]$ of ${\mathbb G}$ such that $A[l] = true$  $\forall l : i \leq l \leq j $.

\section{Computational results}

Table~\ref{hg38-3} reports the data for human chromosome 1 up to $k=40$. 
It appears remarkable that already in the range [28-40] for the $k$ values, the genomic coverage of the $k$-segmentation is almost total (over positions where there is no N), and almost all the $k$-mers present in the chromosome are involved by $k$-spectral maximal segments (see the normalized cardinality of set $U_k$), that is, they have the property to be univocally elongated in the $k$-spectrum. Essentially, we see that for $k=27$ human chromosome~1 may be  covered by 4 millions of relatively short (80 bp long) spectral maximal segments, which have of course a high value of multiplicity in the segmentation, whereas for $k=44$ about half of the spectral maximal segments (1.985.225) of average length 157 cover the chromosome. We observe that he total coverage is reached for $k=216$ with 47.237 maximal segments of average length 5.093 (and maximal length 1.978.600). In this context it is worth to notice that the unique reconstruction of the genome is proved to be possible for $k= mrl+2$, which corresponds to the value 29.265 for human chromosome 1. %Then, in general, over real genomes, where evolution has produced information redundancy by duplication of long genomic portions (then mrl has high values), the $k$-spectral segmentation of sequences appear advantageous as a compact representation of genomes. 

A $k$-segmentation was computed also for {\it Saccharomyces cerevisiae} genome (around 12 Millions bp with 16 chromosomes) where no N characters are present. In all chromosomes, 
we have verified that for $k=mrl +2$, only one spectral segment was found covering the whole chromosome, while for $k$ close to 60, in almost all the cases, spectral segments are less than 100. Table~\ref{cerv} reports the data for chromosome IV (which is the longest one) of {\it Saccharomyces cerevisiae} (with $mrl$ equal to 3.573) up to $k=40$. We remarkably notice that $k=21$ is sufficient to entirely cover the genomic sequence (1.531.933 long) by a segmentation of 1.584 maximal segments (in average shorter than one thousand bp), and that $k=16$ corresponds to a 16-spectral segmentation with a very good coverage, which involves most of the 16-mers present in the genome, into only 8.958 maximal segments having an average length equal to 185,8.

The  $k$-segmentations were also computed, by the previous algorithm, for chromosomes  2, 10, 22, X, Y of Human genome (Hg 38), for $k$-values ranging from 10 to more than 500, where segments are measured (in number and lengths), Tables \ref{hg38-chr2}, \ref{hg38-chr10}, \ref{hg38-chr22}, \ref{hg38-chrx}, \ref{hg38-chry} report the obtained values up to $k= 40$. The results are uniform for all considered chromosomes.

\begin{table}
\begin{tabular}{llllllll}
\hline
\hline
$k$&	$D_k$&	$U_k$&	$U_k / D_k$&	\it Coverage  &\it Maximals&\it AvgL &\it MaxL \notag  \\
10&	1.045.927&	38&			0,0000&	2,25615E-06&	52&			11,00&	11\\
11&	4.046.584&	13.683&		0,0034&	0,000724155&	15.955&		12,08&	17\\
12&	13.963.535&	313.027&		0,0224&	0,014264785&	318.364&		13,21&	24\\
13&	40.048.761&	3.259.722&	0,0814&	0,099454848&	2.645.335&	14,44&	39\\
14&	88.918.637&	15.853.072&	0,1783& 	0,417332951&	10.984.163&	15,66&	54\\
15&	137.433.303&	59.962.772&	0,4363&	0,843980167&	33.214.551&	17,01&	103\\
16&	166.201.314&	119.093.792&	0,7166&	0,921733136&	35.952.622&	19,59&	165\\
17&	179.785.859&	157.738.137&	0,8774&	0,939876978&	22.706.197&	24,43&	375\\
18&	186.272.565&	176.141.059&	0,9456&	0,949814751&	12.897.425&	32,53&	883\\
19&	189.945.022&	184.536.009&	0,9715&	0,957262414&	8.346.098	&	42,45&	1.677\\
20&	192.510.989&	188.930.042&	0,9814&	0,963687855&	6.456.253	&	50,99&	3.903\\
21&	194.595.977&	191.777.370&	0,9855&	0,969077973&	5.624.104	&	57,07&	6.407\\
22&	196.426.475&	193.995.985&	0,9876&	0,973513497&	5.165.406	&	61,69&	14.233\\
23&	198.093.250&	195.897.366&	0,9889&	0,977337135&	4.864.712	&	65,52&	14.887\\
24&	199.637.221&	197.618.157&	0,9899&	0,980668464&	4.623.956	&	69,09&	16.450\\
25&	201.077.240&	199.212.471&	0,9907&	0,98370779&	4.424.312&	72,46&	16.511\\
26&	202.425.001&	200.690.961&	0,9914&	0,986368122&	4.248.537	&	75,75&	21.393\\
27&	203.689.655&	202.070.404&	0,9921&	0,98849874&	4.072.258	&	79,21&	23.140\\
28&	204.878.861&	203.365.399&	0,9926&	0,990262556&	3.907.943&	82,71&	23.289\\
29&	206.000.589&	204.579.712&	0,9931&	0,99181704&	3.746.919&	86,35&	25.316\\
30&	207.060.369&	205.720.228&	0,9935&	0,993210426&	3.601.305	&	89,95&	25.317\\
31&	208.061.649&	206.793.925&	0,9939&	0,994265107&	3.457.401	&	93,71&	26.342\\
32&	209.008.844&	207.810.544&	0,9943&	0,995187959&	3.315.966	&	97,65&	32.156\\
33&	209.904.984&	208.781.444&	0,9946&	0,995923721&	3.169.314	&	101,95&	32.159\\
34&	210.752.552&	209.694.818&	0,9950&	0,996654705&	3.046.401	&	105,98&	32.162\\
35&	211.553.055&	210.557.940&	0,9953&	0,997153583&	2.918.996	&	110,36&	37.175\\
36&	212.308.870&	211.371.979&	0,9956&	0,997652279&	2.795.085	&	114,94&	37.176\\
37&	213.022.579&	212.138.601&	0,9959&	0,997991136&	2.680.028	&	119,56&	37.179\\
38&	213.696.481&	212.864.835&	0,9961&	0,998288267&	2.565.984	&	124,45&	37.180\\
39&	214.332.908&	213.548.776&	0,9963&	0,998509309&	2.457.660	&	129,48&	37.359\\
40&	214.934.436&	214.196.288&	0,9966&	0,99873295&	2.350.398	&	134,82&	37.372\\	
\hline
\end{tabular}
\caption{{\bf Values of spectral segmentation for Hg38 Chr 1}, computed by Procedure 2 and using the IGTools platform~\cite{igtools}. From the left: $k$-column gives the length of $k$-mers; $D_k$-column provides the cardinality of $D_k$ (number of different $k$-mers in the chromosome), $U_k$-column reports the number of $k$-mers occurring in the $k$-spectral maximal segments;  Coverage(\%)-column the percentages of positions of Hg38 Chr 1 covered by the $k$-spectral segmentation; Maximals-column shows the number of $k$-spectral maximal segments, AvgL-column and MaxL-column the average and maximum lengths of  $k$-spectral maximal segments. The length of the chromosome 1 (including the 18.475.410 N characters which were ignored from the collection of $k$-mers) is $248.956.422$~bp.}
\label{hg38-3}
\end{table}

%In the following tables, the term ``Maximals''  stands for ``Maximal segments''.
%~\footnote{This supplementary material is available upon request to the authors.}.  
\begin{table}
\begin{center}
\begin{tabular}{llllllll}
\hline
\hline
$k$&	$D_k$&	$U_k$&	$U_k / D_k$&	\it Coverage  &\it Maximals&\it AvgL &\it MaxL \notag  \\
%k&	$D_k$&		$U_k$&		$U_k / D_k$&	Coverage (\%) & Maximals &Avg.L & MaxL\\
8&	65337&	617& 	0,45915 &	0,3133&	6&	9&	9 \\
9&	239715&	33262&		 0,006695&	0,010416&	1837&	10,037&		12\\
10&	630466&	279271&		0,0920684&	0,2700&		53070&	11,2373&		22\\
11&	1066662&	776349&		0,352781&	0,8457&		231242&	12,7622&		38\\
12&	1328523&	1177684&		0,672753&	0,9934&		288803&	15,2555&		129\\
13&	1434547&	1363931&		0,875117&	0,9994&		166821&	20,8055&		148\\
14&	1470635&	1430564&		0,958397&	0,9998&		66411&	35,8881&		439\\
15&	1482371&	1452182&		0,986831&	0,9999&		23344&	79,4844&		876\\
16&	1486575&	1459635& 	0,99548&		0,9999&		8958		&185,843&	2440\\
17&	1488399&	1462631&		0,998181&	0,9999&		4356&	367,469&		6536\\
18&	1489415&	1464150&		0,999085&	0,9999&		2718	&	580,377&		11110\\
19&	1490140	&1465161		&0,999397	&0,9999		&2061	&761,024		&25903\\
20&	1490735	&1465965	&	0,999531		&0,9999		&1785	&876,975		&65133\\
21&	1491246	&1466635	&	0,999597		&1&			1584		&986,89		&69154\\
22&	1491694	&1467215	&	0,999663		&1&			1421		&1098,87&	69155\\
23&	1492103	&1467741		&0,999694	&1&			1307&	1193,91&		85413\\
24&	1492492	&1468250	&	0,999705		&1&			1213&	1285,75&		85414\\
25&	1492849	&1468714	&	0,99974		&1&			1118		&1394,09&	85417\\
26&	1493188	&1469161		&0,999752&	1&			1059		&1471,44&	85420\\
27&	1493510	&1469594		&0,999763&	1&			1014		&1536,66&	85455\\
28&	1493813	&1470004		&0,999783	&1			&967&	1611,12&		95170\\
29&	1494104	&1470401		&0,999793	&1&			925&		1684,07&		118445\\
30&	1494380	&1470780		&0,999803&	1&			877&		1775,72&		118446\\
31&	1494639	&1471130		&0,999813	&1&			829&		1877,86&		118447\\
32&	1494885	&1471461		& 0,999829&	1&			790&		1970,09&		118448\\
33&	1495122	&1471781		&0,999834&	1 &			767&		2029,24&		118449\\
34&	1495349	&1472088		&0,999843	&1&			729&		2134,35&		146444\\
35&	1495571	&1472389		&0,999846&	1&			711&		2188,55		&146499\\
36&	1495786	&1472682		&0,999849	&1&			689&		2258,35&		154915\\
37&	1495992	&1472961		& 0,99986		&1&			653&		2381,92&		154916\\
38&	1496194	&1473234		&0,999866&	1&			639&		2434,32&		154917\\
39&	1496396	&1473507		&0,999859	&1&			639&		2435,32&		154918\\
40&	1496593	&1473775		&0,999862	&1&			617&		2521,8		&154919\\
\hline
\end{tabular}
\caption{{\bf Values of spectral segmentation for Saccharomyces cerevisiae Chr IV}, computed by Procedure 2 and using the IGTools platform~\cite{igtools}. Columns are as in Table \ref{hg38-3}. The length of the chromosome (where no N characters appear) is $1531933$~bp.}
\label{cerv}
\end{center}
\end{table}

\begin{table}
\begin{center}
\centering
\begin{tabular}{llllllll}
\hline
\hline
$k$&	$D_k$&	$U_k$&	$U_k / D_k$&	\it Coverage  &\it Maximals&\it AvgL &\it MaxL \notag  \\
10	&	1.045.356	&	62	&	0,0001	&	0,000003613	&	79	&	11	&	11	\\
11	&	4.037.486	&	15.091	&	0,0037	&	0,000815795	&	17.279	&	12	&	16	\\
12	&	13.970.306	&	308.947	&	0,0221	&	0,013846816	&	301.339	&	13	&	25	\\
13	&	40.326.969	&	3.210.596	&	0,0796	&	0,097668722	&	2.559.598	&	14	&	38	\\
14	&	91.475.583	&	15.421.454	&	0,1686	&	0,399774431	&	10.398.399	&	16	&	54	\\
15	&	145.002.519	&	59.930.671	&	0,4133	&	0,846513054	&	33.438.889	&	17	&	109	\\
16	&	178.168.268	&	123.940.839	&	0,6956	&	0,934553290	&	38.598.600	&	19	&	188	\\
17	&	194.126.835	&	168.126.622	&	0,8661	&	0,951892994	&	25.112.141	&	24	&	338	\\
18	&	201.601.752	&	189.796.905	&	0,9414	&	0,960309510	&	14.035.702	&	32	&	829	\\
19	&	205.618.151	&	199.614.013	&	0,9708	&	0,966795324	&	8.619.987	&	43	&	1.778	\\
20	&	208.273.542	&	204.520.815	&	0,9820	&	0,971930785	&	6.328.018	&	54	&	3.767	\\
21	&	210.348.652	&	207.521.297	&	0,9866	&	0,976231049	&	5.322.938	&	62	&	6.466	\\
22	&	212.126.808	&	209.757.547	&	0,9888	&	0,980012274	&	4.799.273	&	67	&	11.899	\\
23	&	213.720.398	&	211.619.396	&	0,9902	&	0,983189018	&	4.464.885	&	72	&	14.695	\\
24	&	215.179.675	&	213.269.242	&	0,9911	&	0,985755501	&	4.202.799	&	77	&	17.310	\\
25	&	216.525.819	&	214.780.861	&	0,9919	&	0,987908009	&	3.980.457	&	81	&	21.586	\\
26	&	217.774.255	&	216.168.103	&	0,9926	&	0,989823405	&	3.790.270	&	85	&	31.139	\\
27	&	218.936.692	&	217.448.199	&	0,9932	&	0,991326612	&	3.606.846	&	89	&	31.144	\\
28	&	220.022.185	&	218.640.791	&	0,9937	&	0,992583745	&	3.432.766	&	94	&	31.158	\\
29	&	221.039.081	&	219.751.927	&	0,9942	&	0,993689170	&	3.272.001	&	98	&	31.159	\\
30	&	221.993.683	&	220.786.995	&	0,9946	&	0,994613704	&	3.126.091	&	103	&	35.072	\\
31	&	222.890.490	&	221.756.576	&	0,9949	&	0,995382290	&	2.984.030	&	108	&	35.078	\\
32	&	223.733.892	&	222.668.355	&	0,9952	&	0,996121593	&	2.844.942	&	113	&	35.079	\\
33	&	224.527.803	&	223.532.397	&	0,9956	&	0,996735362	&	2.713.682	&	118	&	35.519	\\
34	&	225.274.155	&	224.343.871	&	0,9959	&	0,997212601	&	2.594.990	&	123	&	35.520	\\
35	&	225.976.292	&	225.103.137	&	0,9961	&	0,997588970	&	2.477.302	&	128	&	35.521	\\
36	&	226.635.933	&	225.818.025	&	0,9964	&	0,997944325	&	2.369.341	&	134	&	35.848	\\
37	&	227.256.084	&	226.488.028	&	0,9966	&	0,998234288	&	2.267.537	&	140	&	35.860	\\
38	&	227.839.791	&	227.119.333	&	0,9968	&	0,998574569	&	2.168.482	&	145	&	35.861	\\
39	&	228.389.481	&	227.711.681	&	0,9970	&	0,998894409	&	2.071.522	&	152	&	35.862	\\
40	&	228.906.606	&	228.271.367	&	0,9972	&	0,999011699	&	1.976.489	&	158	&	38.236	\\
\hline
\end{tabular}
\caption{{\bf Values of spectral segmentation for Hg38 Chr 2}}
\label{hg38-chr2}
\end{center}
\end{table}

\begin{table}
\begin{center}
%\small
\centering
\begin{tabular}{llllllll}
\hline
\hline
$k$&	$D_k$&	$U_k$&	$U_k / D_k$&	\it Coverage  &\it Maximals&\it AvgL &\it MaxL \notag \\
10	&	1.041.093	&	286	&	0,0000	&	0,000030241	&	367	&	11	&	12	\\
11	&	3.936.809	&	25.449	&	0,0023	&	0,002252922	&	27.322	&	12	&	18	\\
12	&	12.817.030	&	480.917	&	0,0368	&	0,036765759	&	462.859	&	13	&	31	\\
13	&	34.145.871	&	3.580.802	&	0,1749	&	0,174926481	&	2.628.021	&	15	&	36	\\
14	&	65.646.514	&	17.099.963	&	0,6659	&	0,665857037	&	11.413.628	&	16	&	55	\\
15	&	90.151.885	&	50.398.126	&	0,9116	&	0,911648152	&	22.403.605	&	17	&	121	\\
16	&	102.828.231	&	82.036.963	&	0,9472	&	0,947221269	&	18.212.617	&	21	&	198	\\
17	&	108.607.958	&	99.323.101	&	0,9590	&	0,959014141	&	10.462.638	&	27	&	449	\\
18	&	111.483.717	&	107.084.861	&	0,9669	&	0,966874359	&	6.111.992	&	36	&	1.546	\\
19	&	113.244.161	&	110.709.583	&	0,9728	&	0,972821263	&	4.250.781	&	46	&	2.204	\\
20	&	114.553.598	&	112.755.264	&	0,9784	&	0,978430121	&	3.489.199	&	54	&	4.035	\\
21	&	115.654.605	&	114.180.258	&	0,9824	&	0,982428689	&	3.125.739	&	59	&	10.768	\\
22	&	116.633.603	&	115.341.284	&	0,9858	&	0,985826249	&	2.918.936	&	63	&	14.196	\\
23	&	117.528.956	&	116.353.857	&	0,9885	&	0,988479620	&	2.762.778	&	67	&	18.967	\\
24	&	118.357.515	&	117.277.400	&	0,9908	&	0,990840418	&	2.634.631	&	70	&	19.354	\\
25	&	119.126.857	&	118.133.609	&	0,9928	&	0,992771187	&	2.519.577	&	74	&	20.823	\\
26	&	119.843.613	&	118.923.757	&	0,9943	&	0,994315060	&	2.420.266	&	77	&	20.825	\\
27	&	120.512.866	&	119.658.553	&	0,9955	&	0,995469199	&	2.318.595	&	81	&	25.320	\\
28	&	121.139.473	&	120.344.058	&	0,9965	&	0,996514958	&	2.223.277	&	84	&	25.322	\\
29	&	121.726.829	&	120.985.037	&	0,9973	&	0,997292661	&	2.131.640	&	88	&	33.768	\\
30	&	122.278.403	&	121.582.101	&	0,9980	&	0,997975957	&	2.049.112	&	92	&	33.771	\\
31	&	122.796.035	&	122.142.297	&	0,9985	&	0,998514103	&	1.962.984	&	96	&	33.773	\\
32	&	123.282.548	&	122.667.660	&	0,9989	&	0,998881107	&	1.886.814	&	99	&	33.775	\\
33	&	123.739.101	&	123.167.264	&	0,9992	&	0,999177739	&	1.806.606	&	104	&	33.777	\\
34	&	124.166.975	&	123.634.195	&	0,9994	&	0,999430442	&	1.730.945	&	108	&	33.779	\\
35	&	124.568.421	&	124.070.103	&	0,9996	&	0,999605749	&	1.660.329	&	112	&	33.781	\\
36	&	124.944.815	&	124.478.890	&	0,9997	&	0,999722136	&	1.590.536	&	117	&	33.783	\\
37	&	125.297.872	&	124.860.960	&	0,9998	&	0,999798684	&	1.524.071	&	122	&	36.729	\\
38	&	125.629.183	&	125.221.245	&	0,9999	&	0,999872643	&	1.456.539	&	127	&	38.370	\\
39	&	125.940.266	&	125.557.286	&	0,9999	&	0,999924443	&	1.398.753	&	132	&	60.547	\\
40	&	126.231.729	&	125.873.920	&	1,0000	&	0,999959411	&	1.341.799	&	137	&	60.549	\\
\hline
\end{tabular}
\caption{{\bf Values of spectral segmentation for Hg38 Chr 10}}
\label{hg38-chr10}
\end{center}
\end{table}

\begin{table}
\begin{center}
%\small
\centering
\begin{tabular}{llllllll}
\hline
\hline
$k$&	$D_k$&	$U_k$&	$U_k / D_k$&\it	Coverage (\%) & \it Maximals&\it AvgL &\it MaxL \notag  \\
9	&	261.825	&	5	&	0,0000	&	0,000001788	&	7	&	10	&	10	\\
10	&	1.019.673	&	1.999	&	0,0020	&	0,000756797	&	2.756	&	11	&	14	\\
11	&	3.506.660	&	79.689	&	0,0227	&	0,023291501	&	86.240	&	12	&	20	\\
12	&	9.548.425	&	866.147	&	0,0907	&	0,172177385	&	735.492	&	13	&	29	\\
13	&	18.068.493	&	4.891.914	&	0,2707	&	0,650453372	&	3.467.102	&	15	&	46	\\
14	&	24.506.912	&	14.053.286	&	0,5734	&	0,885121409	&	6.352.781	&	16	&	82	\\
15	&	27.801.432	&	22.415.019	&	0,8063	&	0,922426014	&	5.120.141	&	20	&	141	\\
16	&	29.318.528	&	26.906.274	&	0,9177	&	0,938427867	&	3.034.727	&	26	&	314	\\
17	&	30.110.958	&	28.924.250	&	0,9606	&	0,949208214	&	1.883.877	&	34	&	668	\\
18	&	30.636.017	&	29.900.266	&	0,9760	&	0,958403696	&	1.419.137	&	41	&	1.685	\\
19	&	31.056.553	&	30.492.368	&	0,9818	&	0,965737956	&	1.236.907	&	46	&	3.585	\\
20	&	31.426.643	&	30.940.639	&	0,9845	&	0,971911536	&	1.149.500	&	49	&	6.485	\\
21	&	31.766.989	&	31.324.505	&	0,9861	&	0,976692104	&	1.092.487	&	52	&	9.854	\\
22	&	32.085.192	&	31.673.457	&	0,9872	&	0,980757296	&	1.050.098	&	55	&	14.755	\\
23	&	32.385.495	&	31.997.039	&	0,9880	&	0,984112984	&	1.015.327	&	57	&	17.495	\\
24	&	32.669.603	&	32.304.683	&	0,9888	&	0,987362364	&	980.900	&	60	&	19.464	\\
25	&	32.938.199	&	32.596.243	&	0,9896	&	0,990175685	&	946.474	&	62	&	21.031	\\
26	&	33.192.010	&	32.871.397	&	0,9903	&	0,992043162	&	912.734	&	65	&	22.724	\\
27	&	33.431.987	&	33.130.689	&	0,9910	&	0,993624555	&	878.002	&	68	&	29.616	\\
28	&	33.659.147	&	33.375.097	&	0,9916	&	0,994783065	&	845.236	&	71	&	29.617	\\
29	&	33.874.638	&	33.605.890	&	0,9921	&	0,995783148	&	813.740	&	73	&	29.618	\\
30	&	34.078.824	&	33.824.614	&	0,9925	&	0,996856749	&	785.507	&	76	&	33.620	\\
31	&	34.272.239	&	34.030.192	&	0,9929	&	0,997512652	&	758.686	&	79	&	33.621	\\
32	&	34.454.766	&	34.226.804	&	0,9934	&	0,997979738	&	728.923	&	82	&	33.626	\\
33	&	34.626.863	&	34.413.820	&	0,9938	&	0,998461074	&	697.974	&	86	&	33.784	\\
34	&	34.789.125	&	34.589.584	&	0,9943	&	0,998695498	&	670.865	&	89	&	33.830	\\
35	&	34.942.372	&	34.754.380	&	0,9946	&	0,998919529	&	645.406	&	92	&	33.831	\\
36	&	35.086.703	&	34.910.137	&	0,9950	&	0,999043534	&	618.723	&	96	&	33.836	\\
37	&	35.222.644	&	35.056.110	&	0,9953	&	0,999186563	&	595.130	&	100	&	35.249	\\
38	&	35.350.607	&	35.194.803	&	0,9956	&	0,999260210	&	571.318	&	103	&	36.970	\\
39	&	35.471.346	&	35.324.799	&	0,9959	&	0,999324000	&	549.538	&	107	&	36.976	\\
40	&	35.585.086	&	35.447.664	&	0,9961	&	0,999460058	&	528.884	&	111	&	36.985	\\
\hline
\end{tabular}
\caption{{\bf Values of spectral segmentation for Hg38 Chr 22}}
\label{hg38-chr22}
\end{center}
\end{table}

\begin{table}
\begin{center}
%\small
\centering
\begin{tabular}{llllllll}
\hline
\hline
$k$&	$D_k$&	$U_k$&	$U_k / D_k$&\it 	Coverage (\%) &\it Maximals&\it AvgL &\it MaxL \notag \\
10	&	1.037.774	&	491	&	0,0005	&	0,000043888	&	623	&	11	&	13	\\
11	&	3.889.504	&	31.867	&	0,0082	&	0,002465811	&	35.538	&	12	&	18	\\
12	&	12.504.932	&	519.962	&	0,0416	&	0,035696493	&	530.923	&	13	&	26	\\
13	&	33.746.525	&	3.471.711	&	0,1029	&	0,145729870	&	2.581.493	&	15	&	39	\\
14	&	67.004.374	&	16.083.058	&	0,2400	&	0,603250867	&	11.410.193	&	16	&	70	\\
15	&	94.629.309	&	50.176.817	&	0,5302	&	0,887882178	&	24.597.625	&	17	&	121	\\
16	&	110.045.379	&	85.018.232	&	0,7726	&	0,937308793	&	22.040.843	&	20	&	451	\\
17	&	117.654.131	&	105.593.254	&	0,8975	&	0,951868860	&	14.160.125	&	25	&	896	\\
18	&	121.730.602	&	115.579.606	&	0,9495	&	0,960595541	&	9.168.860	&	32	&	1.641	\\
19	&	124.355.508	&	120.620.193	&	0,9700	&	0,967176250	&	6.814.620	&	38	&	2.347	\\
20	&	126.354.205	&	123.628.203	&	0,9784	&	0,972524051	&	5.744.645	&	43	&	4.280	\\
21	&	128.036.423	&	125.792.642	&	0,9825	&	0,976775856	&	5.193.767	&	47	&	7.205	\\
22	&	129.518.262	&	127.571.616	&	0,9850	&	0,980340251	&	4.803.598	&	50	&	9.871	\\
23	&	130.854.734	&	129.111.191	&	0,9867	&	0,983110260	&	4.536.017	&	53	&	14.486	\\
24	&	132.074.054	&	130.491.374	&	0,9880	&	0,985695050	&	4.298.488	&	56	&	19.626	\\
25	&	133.189.081	&	131.756.797	&	0,9892	&	0,987901502	&	4.072.902	&	59	&	20.102	\\
26	&	134.213.903	&	132.905.141	&	0,9902	&	0,989864328	&	3.880.410	&	62	&	21.965	\\
27	&	135.159.044	&	133.956.732	&	0,9911	&	0,991405462	&	3.702.602	&	65	&	21.966	\\
28	&	136.032.999	&	134.928.613	&	0,9919	&	0,992785602	&	3.529.928	&	68	&	21.968	\\
29	&	136.844.664	&	135.822.987	&	0,9925	&	0,993861409	&	3.376.285	&	71	&	21.970	\\
30	&	137.598.873	&	136.651.222	&	0,9931	&	0,994778932	&	3.233.787	&	74	&	22.045	\\
31	&	138.300.893	&	137.420.171	&	0,9936	&	0,995770642	&	3.101.559	&	78	&	22.046	\\
32	&	138.955.736	&	138.134.801	&	0,9941	&	0,996451125	&	2.971.323	&	81	&	30.578	\\
33	&	139.566.748	&	138.806.672	&	0,9946	&	0,996995920	&	2.850.829	&	84	&	30.579	\\
34	&	140.137.310	&	139.430.458	&	0,9950	&	0,997497434	&	2.746.491	&	87	&	30.584	\\
35	&	140.670.461	&	140.011.539	&	0,9953	&	0,997900325	&	2.643.366	&	90	&	39.481	\\
36	&	141.169.081	&	140.554.194	&	0,9956	&	0,998246067	&	2.553.300	&	94	&	39.513	\\
37	&	141.635.366	&	141.061.342	&	0,9959	&	0,998532329	&	2.462.499	&	97	&	39.514	\\
38	&	142.072.342	&	141.536.092	&	0,9962	&	0,998691768	&	2.375.435	&	100	&	39.515	\\
39	&	142.482.391	&	141.980.427	&	0,9965	&	0,998991672	&	2.294.023	&	104	&	39.516	\\
40	&	142.867.125	&	142.397.483	&	0,9967	&	0,999122853	&	2.222.933	&	107	&	39.561	\\
\hline
\end{tabular}
\caption{{\bf Values of spectral segmentation for Hg38 Chr X}}
\label{hg38-chrx}
\end{center}
\end{table}

\begin{table}
\begin{center}
%\small
\centering
\begin{tabular}{llllllll}
\hline
\hline
$k$&	$D_k$&	$U_k$&	$U_k / D_k$&\it	Coverage (\%) &\it Maximals&\it AvgL & \it MaxL \notag\\
9	&	258.851	&	184	&	0,0007	&	0,000094794	&	252	&	10	&	12	\\
10	&	952.028	&	9.178	&	0,0096	&	0,004327837	&	11.235	&	11	&	17	\\
11	&	2.950.639	&	161.600	&	0,0548	&	0,059458998	&	171.130	&	12	&	23	\\
12	&	7.349.099	&	925.005	&	0,1259	&	0,260281537	&	772.716	&	14	&	38	\\
13	&	12.604.575	&	4.350.557	&	0,3452	&	0,797859424	&	3.258.408	&	15	&	69	\\
14	&	16.150.413	&	10.425.123	&	0,6455	&	0,942926688	&	4.617.874	&	17	&	115	\\
15	&	17.999.177	&	15.058.625	&	0,8366	&	0,967826477	&	3.437.514	&	20	&	217	\\
16	&	18.975.234	&	17.499.482	&	0,9222	&	0,977689796	&	2.223.413	&	25	&	713	\\
17	&	19.582.524	&	18.720.914	&	0,9560	&	0,983727378	&	1.584.051	&	31	&	1.591	\\
18	&	20.032.378	&	19.425.139	&	0,9697	&	0,987848742	&	1.278.129	&	36	&	2.341	\\
19	&	20.404.478	&	19.918.053	&	0,9762	&	0,990960416	&	1.116.113	&	40	&	4.899	\\
20	&	20.728.314	&	20.311.145	&	0,9799	&	0,993522138	&	1.010.214	&	43	&	8.835	\\
21	&	21.017.072	&	20.649.604	&	0,9825	&	0,995278940	&	924.063	&	47	&	8.836	\\
22	&	21.276.934	&	20.947.952	&	0,9845	&	0,996511003	&	853.382	&	50	&	14.256	\\
23	&	21.512.015	&	21.215.272	&	0,9862	&	0,997469662	&	788.915	&	54	&	14.290	\\
24	&	21.725.180	&	21.457.857	&	0,9877	&	0,998096577	&	728.442	&	58	&	14.294	\\
25	&	21.918.376	&	21.677.034	&	0,9890	&	0,998614237	&	675.225	&	62	&	14.295	\\
26	&	22.094.064	&	21.876.610	&	0,9902	&	0,999058832	&	626.810	&	66	&	14.296	\\
27	&	22.254.431	&	22.056.528	&	0,9911	&	0,999357298	&	584.288	&	70	&	14.323	\\
28	&	22.400.981	&	22.221.033	&	0,9920	&	0,999647057	&	544.951	&	74	&	22.075	\\
29	&	22.535.244	&	22.371.147	&	0,9927	&	0,999741132	&	507.782	&	79	&	24.908	\\
30	&	22.658.352	&	22.507.948	&	0,9934	&	0,999864471	&	475.866	&	83	&	25.084	\\
31	&	22.771.307	&	22.634.213	&	0,9940	&	0,999906985	&	444.811	&	88	&	25.094	\\
32	&	22.875.207	&	22.749.604	&	0,9945	&	0,999936665	&	416.799	&	93	&	25.096	\\
33	&	22.971.055	&	22.855.636	&	0,9950	&	0,999952716	&	390.904	&	99	&	25.101	\\
34	&	23.059.696	&	22.953.556	&	0,9954	&	0,999964679	&	366.515	&	104	&	25.102	\\
35	&	23.141.808	&	23.043.565	&	0,9958	&	0,999973765	&	344.966	&	110	&	28.304	\\
36	&	23.217.796	&	23.127.687	&	0,9961	&	0,999980163	&	324.812	&	115	&	28.305	\\
37	&	23.288.437	&	23.204.700	&	0,9964	&	0,999986788	&	306.251	&	121	&	28.306	\\
38	&	23.354.165	&	23.276.355	&	0,9967	&	0,999990574	&	287.691	&	128	&	29.584	\\
39	&	23.415.300	&	23.343.464	&	0,9969	&	0,999993261	&	271.426	&	135	&	29.585	\\
40	&	23.472.435	&	23.405.683	&	0,9972	&	0,999995722	&	256.379	&	141	&	29.586	\\
\hline
\end{tabular}
\caption{{\bf Values of spectral segmentation for Hg38 Chr Y}}
\label{hg38-chry}
\end{center}
\end{table}

The algorithm of Procedure 2 was used to compute the data reported in all tables, where the minimal values of $k$ indicate the length of the minimal $k$-mer which could be univocally elongated in the $k$-spectrum. These computational experiments were implemented  within the platform IGTools~\cite{igtools}, specifically designed to develop informational analyses of genomes. Data were globally obtained in less than 10 hours running computations. The algorithm was implemented using a suffix array representation of genomes~\cite{abouelhoda2004replacing} with a complexity linear with respect the size of genomes.

\section{Conclusions}
The notion of distribution and of information source has an immense scientific value and appears everywhere, in an enormous number of different contexts. This ubiquity and variety is a sort of evidence of its fundamental role. In the paper we studied a special case of genomic distribution, based on substring occurrences in genomes. The notions of $k$-spectrum, spectral segments and  segmentations, are showed to be strictly related to the internal organisation of genomes, through experimental results on human chromosomes. Only some computational experiments on real genomes are here reported, however many other related ones may be conceived (for example to study the correlation among the informational indexes discussed in this paper, including those of $k$-spectral segmentations), which could assess the real biological importance of the concepts analysed. Specific experiments could suggest further clues on the deep nature of genomic sequences, and on the internal principles that rule their coherence, their plasticity, their equilibria and their instability, by pushing them toward continuous transformations. 

The topics developed in the paper are mainly of theoretical nature, and the genomic terminology is used for the motivation underlying the investigated arguments. As an example we mention the algorithm of Procedure~1, which may become more interesting in the case that some efficient technology providing directly $k$-spectra, independently from sequencing procedures, should be available in the future.

%We are convinced that such deep principles exist, while no deep comprehension of genomes can be obtained by only empirical and statistical analyses. Genomes are pure biological information, therefore specific  informational entities are the essence of their existence and of their evolution.

\

\end{document}